\documentclass{article}

\usepackage[latin1]{inputenc} 
\usepackage[T1]{fontenc}
\usepackage[english]{babel}
\usepackage{amsthm}
\usepackage{lmodern}
\usepackage{amssymb}
\usepackage{mathrsfs}

\title{Greedy Construction of DNA Codes and New Bounds }
\author{Nabil Bennenni, Kenza Guenda \thanks{%
N. Bennenni and K. Guenda are with the Faculty of Mathematics USTHB,
University of Science and Technology of Algiers, Algeria.
email: nbennenni@usthb.dz, ken.guenda@gmail.com}, T. Aaron Gulliver}

\date{ }
\newtheorem{thm}{Theorem}

\newtheorem{defi}[thm]{Definition}

\newtheorem{prop}[thm]{Proposition}
\newtheorem{pf}[thm]{Proof}
\newtheorem{rem}[thm]{Remark}
\begin{document}
\maketitle
\abstract{In this paper, we construct linear codes over $\mathbb{Z}_4$ with bounded $GC$-content. The codes are obtained using a greedy algorithm over $\mathbb{Z}_4$. Further, upper and lower bounds are derived for the maximum size of DNA codes of length $n$ with constant $GC$-content $w$ and edit distance $d$.}

\textbf{keywords:} DNA codes, $GC$-content, edit distance, upper and lower bounds. 

\section{Introduction}

Deoxyribonucleic acid (DNA) contains the genetic program for the
biological development of life.
DNA is formed by strands linked together and twisted in the shape of a double helix.
Each strand is a sequence of four possible nucleotides, two purines,
adenine $A$ and guanine $G$, and two pyrimidines, thymine $T$ and
cytosine $C$.
The ends of a DNA strand are chemically polar with $5'$ and $3'$ ends, which implies that the strands are oriented.
Hybridization, known as base pairing, occurs when a strand binds to another strand, forming a double strand of DNA.
The strands are linked following the Watson-Crick model.
Every $A$ is linked with a $T$, and every $C$ with a $G$, and vice versa.
We denote the complement of $X$ by $\hat{X}$, i.e., $\hat{A}=T,\hat{T}=A,\hat{G}=C$ and $\hat{C}=G$.
The pairing is done in the opposite direction and the reverse order.
For instance, the Watson-Crick complementary (WCC) strand of $3'-ACTTAGA-5'$ is the strand $5'-TCTAAGT-3'$.

The WCC property of DNA strands is used in DNA computing.
In this case the data is encoded using DNA strands, and molecular biology techniques
are used to simulate arithmetic and logical operations.
The main advantages of this approach are huge memory capacity, massive parallelism, and low power molecular hardware and software.
Other applications make use of the properties of DNA~\cite{shoemaker}.

In this paper, we construct linear codes over $\mathbb{Z}_4$ with bounded $GC$-content.
The codes are obtained using a greedy algorithm over $\mathbb{Z}_4$.
Further, upper and lower bounds on the maximum size of DNA codes of length $n$ with constant $GC$-content $w$ and
edit distance $d$ are given.

The choice of the ring $\mathbb{Z}_4$ comes from the fact that the bounded $GC$-content and bounded edit distance properties
are multiplicative over $\mathbb{Z}_4$.
This is not the case over $\mathbb{F}_4$.
The bounded $GC$-constraint ensures that all codewords have thermodynamic characteristics below some threshold.
This is an important criteria for DNA sequences as it reduces the probability of erroneous cross-hybridization.

In~\cite{chee}, Chee and Ling gave an algorithm to construct DNA codes with large
$GC$-content which are optimal only up to $n=12$.
Bishop et al.~\cite{Free} considered the construction of random codes with fixed $GC$-content
using a probabilistic model.
King \cite{bound1} and Condon et al. \cite{condon} gave several upper and lower bounds on the maximum size of DNA codes of length $n$ with constant $GC$-content $w$ and  Hamming distance $d$.
It is well known that the Hamming distance does not capture the thermodynamic and the combinatorial
properties of DNA strand.
In fact, the edit distance is a much more appropriate metric for designing codes for DNA computing.
Thus, in the second part of this paper upper and lower bounds are derived
for the maximum size of DNA codes of length $n$ with constant $GC$-content $w$ and edit distance $d$.

The remainder of this paper is organized as follows.
In Section 2, some preliminary results are presented.
Section 3 employs a greedy algorithm to obtain DNA codes with bounded $GC$-content, and in Section 4
DNA lexicodes are constructed with bounded edit distance.
Upper and lower bounds on the edit distance are also presented.
In addition, examples of DNA codes with bounded $GC$-content and edit distance are given.

\section{Preliminaries}

The ring $\mathbb{Z}_4$ with element $\{0,1,2,3\}$ is considered here
with addition and multiplication modulo $4$.
It is a finite chain ring with maximal ideal $<2>$ and nilpotency index $2$.
The Hamming weight of a codeword $x$ in $\mathbb{Z}_4^n$ is defined as
$w_H(x)=n_1(x)+n_2(x)+n_3(x)$,
and the Hamming distance $d_H (x,y)$ between two codewords $x$ and $y$ as $w_H(x-y)$.
We define the reverse of $x=(x_0x_1\cdots x_{n-1})$ to be $x^R=(x_{n-1}x_{n-2}\cdots x_{1}x_0)$.

The elements $\{0,1,2,3\}$ of $\mathbb{Z}_4$ are in one to one correspondence with the nucleotide DNA bases
$\{A,T,C,G\}$ by the map $\phi$ such that
$0\rightarrow G$, $2\rightarrow C$, $3 \rightarrow T$ and $1\rightarrow A$.

The complement of the codeword $x=(x_0x_1\cdots x_{n-1})$ is the vector
$x^C=(\hat{x_0}\hat{x_1}\cdots \hat{x_{n-1}})$.
The reverse complement (also called the Watson-Crick complement) is
$x^{RC}=(\hat{x_{n-1}}\hat{x_{n-2}}\cdots \hat{x_{1}} \hat{x_{0}})$.
For $ x \in \mathbb{Z}_4$, $\hat{x}$ is defined to be $\hat{ \phi (x)}$.
A linear code $\mathcal{C}$ is said to satisfy the reverse constraint, respectively the reverse-complement constraint
if for all $x\in \mathcal{C}$ we have $x^{R}\in \mathcal{C}$, respectively $x^{RC}\in \mathcal{C}$.

\subsection{Construction of Lexicodes over $\mathbb{Z}_4$}

The construction of lexicodes over $\mathbb{Z}_4$ given in \cite{GG13} is now reviewed.
A linear code $\mathcal{C}$ of length $n$ over $\mathbb{Z}_4$ is an additive code over $\mathbb{Z}_4^n$.
Thus $\mathbb{Z}_4^n$ is a linear code over $\mathbb{Z}_4$ with basis $B=\{b_1\cdot\cdot\cdot b_n\}$.
With respect to this basis, we recursively define a lexicographically ordered list
$V_i=x_1,x_2,\cdot\cdot\cdot,x_{4^i}$ as follows
\[
V_0:=0
\]
\[
V_i:=V_{i-1},b_i+V_{i-1},2b_i+V_{i-1},3b_i+V_{i-1},1\leq i\leq n.
\]
In this way $|V_i|=4^i$, and ${\mathbb{Z}_4^n}$ can be associated with $V_n$.
Assume now that we have a property $P$ which can test if a vector $c\in \mathbb{Z}_4^n$ is selected or not.
The selection property $P$ on $V$ can be seen as a boolean valued function
\[
P:V\rightarrow \{True,False\},
\]
that depends on one variable.
Over $\mathbb{Z}_4$, the property $P$ is called a multiplicative property if $P[x]$ is true implies $P[3x]$ is true.
The following greedy algorithm provides lexicodes over $\mathbb{Z}_4^n$ \cite{GG13}.

\hfill \\
\noindent{\bf Algorithm 1}
\begin{enumerate}
\item $\mathcal{C}_{0}:=0;i:=1;$
\item select the first vector $a_{i} \in V_{i}\backslash V_{i-1}$ such that $P[2a_{i}+c]$
for  all $c \in \mathcal{C}_{i-1}$;
\item if such an $a_{i}$ exists, then $\mathcal{C}_{i}:=\mathcal{C}_{i-1},a_{i}+\mathcal{C}_{i-1},2a_{i}+\mathcal{C}_{i-1},3a_{i}+\mathcal{C}_{i-1};$

    otherwise $\mathcal{C}_{i}:=\mathcal{C}_{i-1}$;
\item $i:=i+1;$ return to 2.
\end{enumerate}

For $0<i\leq n$, the code $\mathcal{C}_i$ is forced to be linear
because all linear combinations of the selected vectors
$a_{i1},\cdots,a_{il}$, $l\leq i$, are taken.
The code $\mathcal{C}_i$ has a `basis' formed from
$a_{i1},\cdots,a_{il}$, so we have a nested sequence of linear codes
\[
0=\mathcal{C}_0\subseteq \mathcal{C}_1\subseteq \cdot\cdot\cdot\subseteq \mathcal{C}_n.
\]
$\mathcal{C}_n$ is the lexicode and is
denoted $\mathcal{C}_n=\mathcal{C}(B,P)$ where $B$ is the ordering and $P$
is the selection property.
We have the following result.
\begin{thm}(\cite[Theorem 4]{GG13})
\label{lem:2.2}
For any basis $B$ of $R^n$ and any multiplicative
selection criterion $P$, the lexicode $\mathcal{C}(B,P)$ is linear and
$P[x]$ holds for each codeword $x \neq 0$.
\end{thm}

\section{A Greedy Algorithm for Bounded $GC$-content DNA Codes}
In this Section we construct DNA codes with bounded $GC$-content using Algorithm 1.
We begin with the following definition.
\begin{defi}
\label{def:bounded}
Let $\mathcal{C}$ be a linear code over ${\mathbb{Z}_4}^n$.
The $GC$-content of a codeword $x\in \mathcal{C}$, denoted by
$GC(\phi(x))$, is the number of occurrences of $G$ and $C$ in $\phi(x)$
\[
GC(\phi(x))=|\{ 1\le i \le n;\, \phi(x)_i\in \{G,C\}\}|=w_{GC}(\phi(x)).
\]
We say that a subset $\mathcal{C}$ of ${\mathbb{Z}_4^n}$ satisfies the
bounded $GC$-content constraint if there exists a positive integer $w$
such that $GC(\phi(x)) \ge w, \, \forall  \, x \in \mathcal{C}$.
\end{defi}
\begin{rem}
Definition~\ref{def:bounded} differs from the conventional definition~\cite{GG13,GGS}.
The bounded $GC$-content constraint ensures that all codewords have a
hybridization energy below some threshold, which results in stable DNA strands.
\end{rem}
\begin{prop}
The property $P_{1}[x]$ is true if and only if $w_{GC}(\phi(x)) \ge w$ is a
multiplicative property over $\mathbb{Z}_4$.
\end{prop}
\begin{pf}
Let $x \in {\mathbb{Z}_4}^n$ such that $w_{GC}(\phi(x)) \ge w$.
Multiplying the vector $x$ by $3$ does not change the number of $0$'s and $2$'s.
This gives that $w_{GC}(\phi(3x))=w_{GC}(\phi(x)) \ge w$, and the result follows.
\end{pf}

\subsection{Construction Results}

In this section, construction results are presented for linear
codes over $\mathbb{Z}_4$ with bounded $GC$-content.
In this case, the verification step for $w_{GC}(\phi(2x)) \ge w$ in Algorithm 1 can be eliminated.
This is because for $x \in \mathbb{Z}_4^n$, $w_{GC}(\phi(x)) \ge w$ implies that
$w_{GC}(\phi(2x)) \ge w$, and this improves the speed of the algorithm.
Some of these codes attain upper bound (5) given in \cite[Proposition 1]{bound1}.
Furthermore, the codes obtained are linear as opposed to those in \cite{aboul}.
Table 1 gives DNA lexicodes over ${\mathbb{Z}_4^n}$ obtained using the selection property $P_{1}[x]$ ($w_{GC}(\phi(x)) \ge w$).
The DNA code strands corresponding to the first and second codes in Table 1 are
given in Tables 2 and 3, respectively.

\begin{table}
\caption{ DNA Lexicodes over ${\mathbb{Z}_4^n}$ Obtained using the Selection Property $P_{1}[x]$ ($w_{GC}(\phi(x)) \ge w$)}
\begin{center}
\begin{tabular}{|c|c|c|c|c|}
  \hline
  $n$ & $w$ & $d_H$ & Basis of $\mathbb{Z}_4$ & Basis of $\mathcal{C}(B,P)$  \\
  \hline
  8&4&4& Canonical basis & $21111000$ \\
   & & & &$13210100$ \\
   & & & &$32310010$\\
   \hline
     10& 6 & 4 & Canonical basis & $2111100000 $\\
    & & & &$1321010000 $\\
    & & & &$3231001000 $\\
  \hline
  10 & 10 & 1 & Canonical basis &$2000000000$\\
  & & & &$0200000000$\\
  & & & &$0020000000$\\
  & & & &$0002000000$\\
  & & & &$0000200000$\\
  & & & &$0000020000$\\
  & & & &$0000002000$\\
  & & & &$0000000200$\\
  & & & &$0000000020$\\
  & & & &$0000000002$\\
  \hline
  12 & 12 & 1 & Canonical basis &$200000000000$\\
  & & & &$020000000000$\\
  & & & &$002000000000$\\
  & & & &$000200000000$\\
  & & & &$000020000000$\\
  & & & &$000002000000$\\
  & & & &$000000200000$\\
  & & & &$000000020000$\\
  & & & &$000000002000$\\
  & & & &$000000000200$\\
  & & & &$000000000020$\\
  & & & &$000000000002$\\
  \hline
\end{tabular}
\end{center}
\end{table}

\begin{table}
\caption{DNA Code Strands Corresponding to the Linear Code in the
First Row of Table 1}
\begin{center}
\begin{tabular}{|c|c|c|c|}
  \hline
  GGGGGGGG & GGGGCCCC & CCCCGGGG & GAAAACCC\\
  GGCCGGCC & CCGGCCGG & CGCGCGCG & GATTTCCC\\
  GGGCCCGC & GGGCCCCG & GGGAAAAC & AACCCGTT\\
  CAAAAGGG & AAAAGGGC & GGGAAACT & TTAACCCG\\
  TGGGAAAC & CTGGGAAA & CTAAAGGG & CCCAAAGT\\
  GGAAACTG & ACTGGGAA & GAAACTGG & TGGGCTTT\\
  AAACTGGG & AACTGGGA & GGGAACTT & CCCGATTT\\
  TTGGGAAC & ACTTGGGA & TGGGAACT & TTCCCGAA\\
  CTTGGGAA & AACTTGGG & GGGCTTAA & AATTCCCG\\
  GGAACTTG & GAACTTGG & AGGGCTTA & TAAACCCG\\
  AAGGGCTT & TTAAGGGC & CTTAAGGG & TTGGGCTT\\
  GCTTAAGG & TAAGGGCT & GGCTTAAG & GCCCTTTT\\
  ATTGGGCA & TTGGGCAA & GGGACTTT & GACCCTTT\\
  TTTGGGAC & TTGGGACT & TGGGACTT & CAATTCCG\\
  CTTTGGGA & TTTACGGG & GACTTTGG & GAACCCTT\\
  CATTTGGG & GGACTTTG & GGGCTTTT & GAAACCCT\\
  \hline
\end{tabular}
\end{center}
\end{table}
\begin{table}
\caption{ DNA Code Strands Corresponding to the Linear Code in the
Second Row of Table 1}
\begin{center}
\begin{tabular}{|c|c|c|c|}
  \hline
  GGGGGGGGGG & TCTAGGAGGG & GGCCGGCGGG & ACATGGTGGG\\
  ATCAGAGGGG & GAACGAAGGG & TTGTGACGGG & CATGGATGGG\\
  CCGCGCGGGG & AGTTGCAGGG & GCCGGCCGGG & TGAAGCTGGG\\
  TACTGTGGGG & CTAGGTAGGG & AAGCGTCGGG & GTTCGTTGGG\\
  CAAAAGGGGG & ATGCAGAGGG & GATTAGCGGG & TTCGAGTGGG\\
  TGTGAAGGGG & CCCAAAAGGG & AGACAACGGG & GCGTAATGGG\\
  GTATACGGGG & TAGGACAGGG & CTTAACCGGG & AACCACTGGG\\
  ACTGATGGGG & GGCAATAGGG & TCACATCGGG & CGGTATTGGG\\
  GCCCCGGGGG & TGATCGAGGG & CCGGCGCGGG & AGTACGTGGG\\
  AAGTCAGGGG & GTTGCAAGGG & TACACACGGG & CTACCATGGG\\
  CGCGCCGGGG & ACAACCAGGG & CGGCCCCGGG & TCTTCCTGGG\\
  TTGACTGGGG & CATCCTAGGG & ATCTCTCGGG & GAAGCTTGGG\\
  CTTTTGGGGG & AACGTGAGGG & GTAATGCGGG & TAGCGGTGGG\\
  TCAGTAGGGG & CGGATAAGGG & ACTCTACGGG & GGCTTATGGG\\
  GATATCGGGG & TTCCTCAGGG & CAATTCCGGG & ATGGTCTGGG\\
  AGACTTGGGG & GCGTTTAGGG & TGTGTTCGGG & CCCATTTGGG\\
  \hline
\end{tabular}
\end{center}
\end{table}
\section{DNA Codes and Edit Distance}
The edit distance has been used for biological computation, in particular for two types of genetic mutation.
The first is the substitution of nucleotides and consists of two possible mutations:
\begin{itemize}
\item
{\it Transition}: a purine is replaced by a purine $(A \leftrightarrow G)$ or
a pyrimidine is replaced by a pyrimidine $(T \leftrightarrow C)$.\\
{\it Transversion}: a purine is replaced by a pyrimidine or the reverse (eg. $A\leftrightarrow C$).
\item Modification using insertions and deletions.
\end{itemize}
In this section, we consider the edit distance in the greedy algorithm
in order to find large sets of DNA codewords of length $n$ with given $w_{GC}$ and minimum edit distance $d$.
We begin by providing a definition of edit distance which follows the presentation in~\cite{edit}.

Let $\mathcal{A}$ and $\mathcal{B}$ be finite sets of distinct symbols
and let $x^t\in \mathcal{A}^t$ denote an arbitrary string of length $t$ over $\mathcal{A}$.
The string edit distance is characterized by a triple
$<\mathcal{A},\mathcal{B},c>$ consisting of the finite sets
$\mathcal{A}$ and $\mathcal{B}$, and the primitive function
$c:E\rightarrow \mathbb{R}_+$ where $\mathbb{R}_+$ is the set of nonnegative reals,
$E=E_s\cup E_d \cup E_i$ is the set of primitive edit operations,
$E_s=\mathcal{A}*\mathcal{B}$ is the set of substitutions,
$E_d=\mathcal{A}*{E}$ is the set of deletions, and
$E_i=E\times\mathcal{B}$ is the set of insertions.
Each triple $<\mathcal{A},\mathcal{B},c>$ induces a distance
function $d_c:\mathcal{A}^{*}\times\mathcal{B}^*\rightarrow
\mathbb{R}_+$ that maps a string $x^t$ to a nonnegative value \cite{edit}.
\begin{defi}
The edit distance $d_c(x^t,y^v)$ between two strings
$x^t\in \mathcal{A}^t$ and $y^v\in \mathcal{B}^v$ is
defined recursively as
\[
d_c(x^t,y^v)=\min\left\{
                   \begin{array}{l}
                     c(x^t,y^v)+d_c(x^{t-1},y^{v-1}), \\
                     c(x^t,\epsilon)+d_c(x^{t-1},y^v), \\
                     c(\epsilon,y^v)+d_c(x^t,y^{v-1});
                   \end{array}
                 \right.
\]
where $d_c(\epsilon,\epsilon)=0$ and $\epsilon$ denotes the empty string of length $n$.
\end{defi}
The edit distance constraint for a DNA code $\mathcal{C}$
is $d_c(x,y)\geq d \forall x,y \in \mathcal{C}$,
$x\neq y$, for some prescribed minimum edit distance $d$.
The edit distance constraint can reduce non-specific hybridization between distinct codewords,
as well as allow for the correction of insertion, deletion and substitution errors in codewords.
\newline
\begin{prop}
The property $P_2[x]$ is true only if $d_c(\phi(x),\phi(y))\leq w$ is a multiplicative property over
$\mathbb{Z}_4$.
\end{prop}
\pf
Let $x\in{\mathbb{Z}_4}^n$  and $y\in{\mathbb{Z}_4}^n$.
Multiplying $x$ by $3$ and $y$ by $3$ does not change the number of $0$'s and $2$'s.
Therefore the number of $1$'s and $3$'s also does not change, so
\[
n_1(x)+n_0(x)+n_2(x)+n_3(x) = n_1(3x)+n_0(3x)+n_2(3x)+n_3(3x).
\]
This also holds for $y$ and thus $d_c(x,y)=d_c(3x,3y)$.
\qed


Now we use Algorithm 1 to construct linear codes over $\mathbb{Z}_4$
with GC-content bounded by $w$ and edit distance $d_c(\phi(x),\phi(y))$ such that $x\in {\mathbb{Z}_4^*}$ and $y\in {\mathbb{Z}_4^*}$.
The results are given in Table 4.

\begin{table}
\caption{DNA Lexicodes over ${\mathbb{Z}_4}^n$ Obtained using the Selection Property $P_{2}[x]$ ($d_c(\phi(x),\phi(y)) \leq m$)}
\begin{center}
\begin{tabular}{|c|c|c|c|c|c|}
  \hline
  $n$ &$\phi(x)$&$m$ & $w_{GC}$ & Basis of $\mathbb{Z}_4$ & Basis of $\mathcal{C}(B,P)$  \\
  \hline
  4 & GGGG& 1 &  4 & Canonical basis & 2222  \\
  & & &  & &  2202 \\
   & & &  & &  2220 \\
   & & &  & &  2022 \\
    \hline
  4 & GCGC& 2 & 4 & Canonical basis & 2020 \\
   & & &  & & 0022 \\
   & & &  & & 0220 \\
   & & & & & 2222 \\
\hline
\end{tabular}
\end{center}
\end{table}
\subsection{Upper and Lower Bounds}

Let $A_4(n,d)$ be the maximum size of a code over $\mathbb{Z}_4$ with length $n$ and minimum edit distance $d$.
Let $A_4^{GC}(n,d,w)$ be the maximum size of a DNA code with length $n$, minimum edit distance $d$,
and fixed GC weight $w$.
Further, let $A_4^{R,GC}(n,d,w)$, respectively $A_4^{RC,GC}(n,d,w)$ be the maximum size of a DNA code with length $n$, minimum edit distance $d$, and
fixed GC weight $w$, that satisfies the reverse constraint, respectively the reverse-complement constraint.
The purpose of this section is to give upper and lower bounds on these quantities.
We have the following theorem.
\begin{thm}
For $n>0$ with $0\leq d\leq n$ and $0\leq w\leq n$, the following results hold.
\begin{equation}
 {A_{4}}^{GC}(n,d,0)=A_2(n,d),\\
\end{equation}

\begin{equation}
 {A_4}^{GC}(n,d,w)=A_4^{GC}(n,d,n-w),
\end{equation}
and if $w=n/2$ then
\begin{equation}
 A_4^{GC}(n,d,w)=4.
\end{equation}
\end{thm}
\begin{pf}
The analogous result for DNA codes with $GC$-content and Hamming distance was given in \cite{bound1}.
The corresponding proof is employed here for the edit distance.\\
(1): Let $\mathcal{C}$ be a linear code over $\mathbb{Z}_{4}^{n}$ with $w_{GC}(\phi(\mathcal{C}))=0$.
Then $\mathcal{C}$ contains only $0$'s and $1$'s,
so $\mathcal{C}$ can be considered as a binary code which gives ${A_{4}}^{GC}(n,d,0)=A_2(n,d)$.\\
(2): Since $w_{GC}(\phi(\mathcal{C}))=n-w_{AT}(\phi(\mathcal{C}))$, interchanging the $A$'s with $C$'s and $T$'s with
$G$'s gives $w_{GC}(\phi(\mathcal{C})) =n-w$, so that $A_4^{GC}(n,d,w)=A_4^{GC}(n,d,n-w)$.\\
(3): Since $A_4^{RC,GC}(n,d,w)\leq A_4^{GC}(n,d,w)$, by \cite[Theorem 5]{bound} we have that\\
$A_4^{RC,GC}(n,d,w) = 2$.
Then $4\leq A_4^{GC}(n,d,w)$, and by the pigeonhole principle\\
$A_4^{GC}(n,d,w)\geq 4$, so that $A_4^{GC}(n,d,w)=4$.
\end{pf}

We have the following relationship between the $GC$-content of a code and the code size over the alphabet $\{A,T,C,G\}$.
\begin{prop}
\begin{equation}
 A_4^{GC}(n,d,w)\geq A_4^{GC}(n+1,d+1,w).
\end{equation}
\begin{equation}
 A_4^{GC}(n,d,w)\geq A_4^{GC}(n+1,d,w)/4.
\end{equation}
\end{prop}
\begin{pf}
The analogous result for DNA codes with unrestricted $GC$-content and Hamming distance was given in \cite{condon}.
The corresponding proof is employed here for the edit distance.
\newline
(4): A $(n,A_4^{GC}(n+1,d+1,w),d,w)$ code can be obtained from a $(n+1,A_4^{GC}(n+1,d+1,w),d+1,w)$
code by removing a symbol from each codeword such that their $GC$-content is preserved.
\newline
(5): If all the codewords in a $(n+1,A_4^{GC}(n+1,d,w),d,w)$ code are partitioned
into four subsets according to the first symbol, one of the subsets will have size at least $A_4^{GC}(n+1,d,w)/4$ and thus is a $(n+1,A_4^{GC+}(n+1,d,w)/4,d,w)$ code.
By removing the (common) symbol from all codewords in the largest subset, a $(n,A_4^{GC+}(n+1,d,w)/4,d,w)$ code is obtained.
\end{pf}

We have the following relationship between the $GC$-content of a reverse code and the code size over the alphabet $\{A,T,C,G\}$.
\begin{prop}
\begin{equation}
 A_4^{GC,R}(n-1,d,w) \leq A_4^{GC,R}(n,d,w) \leq A_4^{GC,R}(n,d-1,w).
\end{equation}
\begin{equation}
  A_4^{GC,R}(n-1,d,w)\geq A_4^{GC,R}(n,d,w)/4.
\end{equation}
\end{prop}
\begin{pf}
The analogous result for DNA codes with unrestricted $GC$-content and Hamming distance was given in \cite{condon}.
The corresponding proof is used here for the edit distance.
\newline
(6): By the construction of codes over ${\mathbb{Z}_4}$, we obtain $4^n$ codewords of length $n$ and $4^{n-1}$ codewords of length $n-1$,
and the result follows.
\newline
(7): The codewords of a $\mathcal{C}(n,A_4^{GC,R}(n,d,w),d)-$code over ${\mathbb{Z}_4}$
can be partitioned into four subsets denoted $C_1,C_2,C_3,C_4$ such that the size of subset $C_1$ is at least $ A_4^{GC,R}(n,d,w)/4$ and
$C_1$ is a $(n,A_4^{GC,R}(n,d,w)/4,d)$ code.
Removing a symbol from the codewords of $C_1$ such that the distance $d$ and weight $w$ are maintained,
we obtain a $(n-1,A_4^{GC,R}(n,d,w),d)$ code, and the result follows.
\end{pf}

\begin{prop}
For $0\leq d \leq n$ and $0\leq w \leq n$
\[
A_4^{GC,RC}(n,d,w)=A_4^{GC,R}(n,d,w),
\]
if $n$ is even, and
\[
A_4^{GC,R}(n,d+1,w) \leq A_4^{GC,RC}(n,d,w)\leq A_4^{GC,R}(n,d-,w),
\]
if $n$ is odd.
\end{prop}
\begin{pf}
The analogous result for DNA codes with unrestricted $GC$-content and edit distance was given in \cite{bound1}.
The corresponding proof is employed here for the edit distance.
Given a set of codewords of length $n$, if we replace all entries in any subset of the positions by their complement,
the $GC$-content of these codewords is preserved, as well as the edit distance between any pair of codewords.
The edit distance between a codeword and the reverse or reverse-complement of the other codewords is not in general preserved,
but if $n$ is even and the first $n/2$ coordinates of each codeword $x_i$ are replaced by their complements to form a new codeword $y_i$,
then $d_c(x_i,x^R)=d_c(y_i,y_j^{RC})$ for all codewords $x_i$ and $x_j$.
Similarly, if $n$ is odd and the first $(n-1)/2$ coordinates of each codeword $x_i$ are replaced by their complements to form $y_i$, then $|d_c(x_i,x_j^R)-d_c(y_i,y_j^{RC})|\leq 1$.
\end{pf}

\end{document}